\documentstyle[prl,aps,epsf]{revtex}
\begin{document}
\draft
\twocolumn[\hsize\textwidth\columnwidth\hsize\csname
@twocolumnfalse\endcsname
\title{Fractal Networks, Braiding Channels, 
and Voltage Noise in Intermittently
Flowing Rivers of Quantized Magnetic Flux}
\author{C.~J.~Olson, C.~Reichhardt, and Franco Nori}
\address{Department of Physics, The University of Michigan, Ann
Arbor, Michigan 48109-1120}
\date{\today}
\maketitle
\begin{abstract}

\vspace{-10pt}
We analyze the microscopic
dynamics of vortex motion through channels that form river-like
fractal networks in a variety of superconducting samples,
and relate it to macroscopic measurable quantities
such as the power spectrum.
As a function of pinning strength, we calculate 
the fractal dimension, tortuosity, and the corresponding 
voltage noise spectrum.
Above a certain pinning strength, a remarkable universal drop in
both tortuosity and noise power occurs when the 
vortex motion changes from shifting braiding channels to unbraided channels.
We compare our results with experiments.
\end{abstract}
\vspace{-3pt}
\pacs{PACS numbers: 74.60.Ge, 62.20.Fe }
\vspace{-20 pt}
\vskip2pc]

{\it Introduction.\/}---The complex dynamics of a 
moving superconducting vortex lattice interacting with material defects has 
attracted considerable experimental and theoretical attention, 
with the observation of intricate channels of vortex motion 
both in simulations \cite{Ref1,Ref2,Ref3-4}, 
beginning with the seminal work in Ref.~\cite{Ref5},
and through Lorentz microscopy \cite{Ref6}.
Similar channel structures have been observed in 
a large variety of systems, including fluid flow in a disordered landscape, 
Josephson junctions, Wigner crystals, magnetic bubbles, and
stress networks in granular systems.
These channels resemble the fractal basins created by natural rivers 
\cite{Ref7} and other fractal network systems (e.g., percolation).

A quantitative microscopic understanding of the characteristics of the 
channels and their effect on macroscopic measurements
is particularly important for superconducting
systems, in which the disorder can be controlled.
Different strengths of disorder produce very different
flow patterns, ranging from elastic flow to plastic flow \cite{Ref1},
characterized by vortices that either remain pinned 
or move intermittently through a vortex river.  
Since these flows can be inferred experimentally via
the voltage noise produced by the moving vortices 
\cite{Ref8,Ref9,Ref10,Ref11},
a deeper understanding of the relationship between the noise
characteristics and the properties of the
vortex channels would lend insight into the experimental systems.

The complex vortex channel network \cite{Ref12} 
observed in some regimes is
difficult to treat analytically.  Until now, the channel structure has been 
studied qualitatively in simulations \cite{Ref1,Ref2,Ref3-4,Ref5},
and only transitions
caused by changes in driving force have been considered 
(e.g., Refs.~\cite{Ref3-4}).
Transitions in driven vortex lattices caused by different 
disorder strength (e.g., Ref.~\cite{Ref2})
are more difficult to study since a separate simulation is
required for each disorder strength.
We use a large-scale parallel simulation to probe 21 samples 
spanning an order of magnitude of pinning strengths, and present
a systematic study of the transition from one plastic flow phase
to another.
We quantify the fractal nature 
and the tortuosity of the vortex 
channels in the plastic flow for the first time, 
and show how both evolve with disorder strength.
We observe remarkable correlations between
microscopic quantities such as the tortuosity and macroscopic
measures such as voltage noise power,
corresponding to changes in the microscopic nature of the channel
flow.

{\it Simulation.\/}---We model a transverse two-dimensional slice
(in the $x$--$y$ plane) of a $T=0$
zero-field-cooled superconducting infinite slab containing
rigid vortices that are parallel
to the sample edge (${\bf H}=H{\hat {\bf z}}$).
Vortices nucleate along one edge of the sample at regular time intervals,
enter the superconducting slab under the force of
their mutual repulsion, pass through
a pinned region $36\lambda \times 36\lambda$ in size (where $\lambda$ is
the penetration depth) where a flux gradient naturally forms \cite{Ref13},
and are removed at the other sample edge.
Up to 1000 vortices are simultaneously inside the sample, which is 
periodic only in the $y$ direction transverse to the gradient.  

The vortex-vortex repulsion is correctly represented
by a modified Bessel function, $K_{1}(r/\lambda)$.
The vortices also interact with 1943 non-overlapping
attractive parabolic wells of radius $\xi_{p}=0.15\lambda$,
representing a density of pinning sites 
$n_{p} = 1.0/\lambda^{2}$.
The maximum pinning force, $f_{p}$, of wells in a given sample has
a Gaussian distribution. We consider 21 samples 
(a much larger number of parameters than in typical simulations \cite{Ref5}) 
with mean values of $f_{p}$ ranging from 
$f_{p} = 0.06 f_{0}$ to $f_{p} = 1.0 f_{0}$,  
where $f_{0} = \Phi_{0}^{2}/8\pi^{2}\lambda^{3}$.

The overdamped equation
of vortex motion is
${\bf f}_{i}={\bf f}_{i}^{vv} + {\bf f}_{i}^{vp}=\eta{\bf v}_{i} \ ,$
where the total force ${\bf f}_{i}$ on vortex $i$ (due to other vortices
${\bf f}_{i}^{vv}$, and pinning sites ${\bf f}_{i}^{vp}$) is
given by
${\bf f}_{i} = \ \sum_{j=1}^{N_{v}}\, f_{0} \,\ K_{1} \hspace{-2pt}
\left( |{\bf r}_{i} - {\bf r}_{j} | /\lambda \right)
\, {\bf {\hat r}}_{ij} 
+  \sum_{k=1}^{N_{p}} (f_{p}/\xi_{p}) \
|{\bf r}_{i} - {\bf r}_{k}^{(p)}| \ \ \Theta \hspace{-2pt} \left[
(\xi_{p} - |{\bf r}_{i} - {\bf r}_{k}^{(p)} |)/\lambda \right] \
{\bf {\hat r}}_{ik} $.
Here, $ \Theta$ is the Heaviside step function,
${\bf r}_{i}$ (${\bf v}_{i}$) is the location (velocity) of the $i$th vortex,
${\bf r}_{k}^{(p)}$ is the location of the $k$th pinning site,
$\xi_{p}$ is the pinning site radius,
$N_{p}$ ($N_{v}$) is the number of pinning sites (vortices),
${\bf {\hat r}}_{ij}=({\bf r}_{i}-{\bf r}_{j} )/
|{\bf r}_{i} -{\bf r}_{j}|$,
${\bf {\hat r}}_{ik}=({\bf r}_{i}-{\bf r}_{k}^{(p)})/
|{\bf r}_{i}-{\bf r}^{(p)}_{k}|$,
and we take $\eta=1$.
We measure all forces in units of
$f_{0}=\Phi_{0}^{2}/8\pi^{2}\lambda^{3}$
and lengths in units of the penetration depth $\lambda$.
We run a highly optimized parallel code on IBM SP parallel computers  
to carefully characterize a wide range of parameters, and
we equilibrate each sample for at least
$10^{6}$ MD steps before taking high resolution data. 
Due to the open boundary conditions in the $x$ direction,
the ratio of vortices to pins, $n_{v}/n_{p}$, is not directly controlled.
Instead, it emerges naturally as the system equilibrates.
Further simulation details appear in Ref.~\cite{Ref2}.

\begin{figure}
\centerline{
\epsfxsize=3.3in
\epsfbox{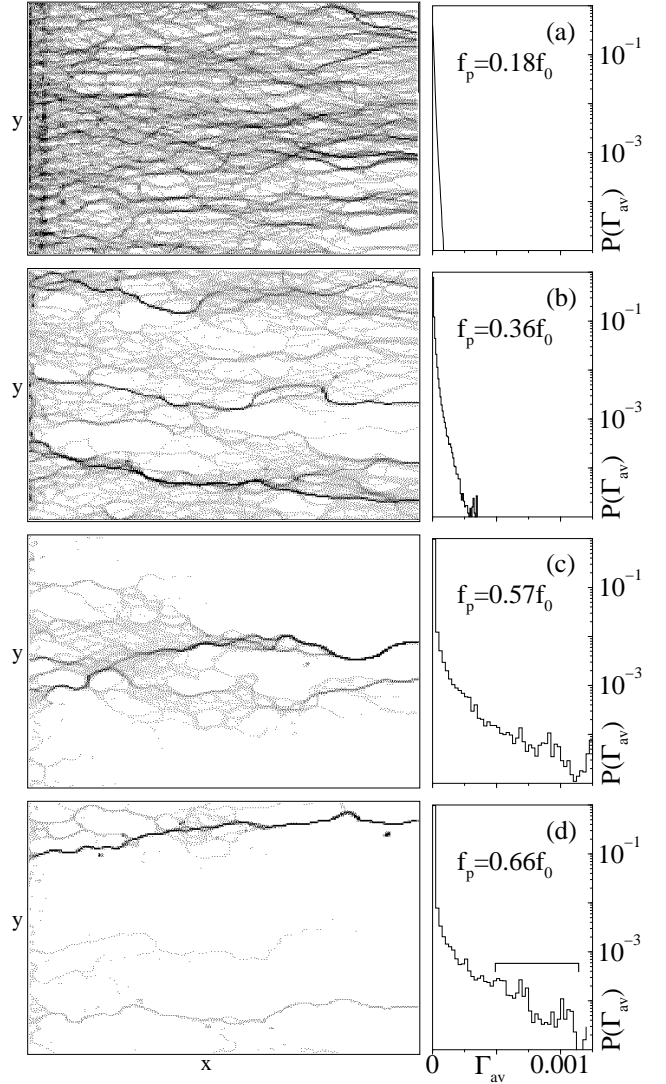}}
\caption{Right panels: 
top view of 
samples with four different pinning strengths:
(a) $f_{p} = 0.18 f_{0}$, (b) $f_{p} = 0.36 f_{0}$, 
(c) $f_{p} = 0.57 f_{0}$, (d) $f_{p} = 0.66 f_{0}$.
The channels most often followed by vortices are darker.
The number of channels decreases as the pinning strength $f_{p}$ increases.
Left panels: distribution of the average rate 
$\Gamma_{\rm av}$ at which vortices pass through
individual grid points in the river.
For strong pinning, the few remaining channels are frequently traveled
[see brace in (d)].
}
\label{fig:1}
\end{figure}

\begin{figure}
\centerline{
\epsfxsize=3.4in
\epsfbox{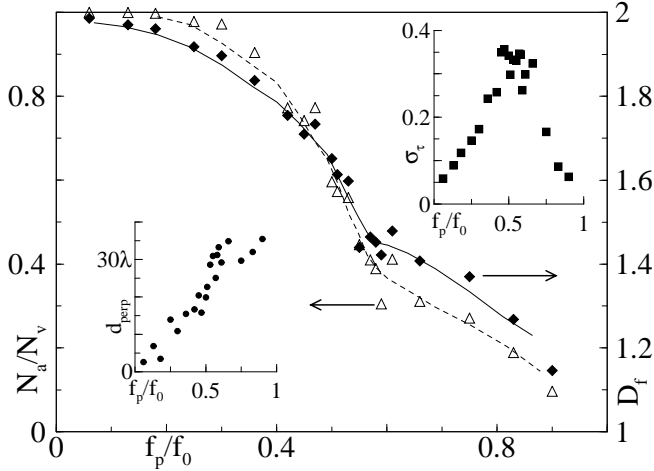}}
\caption{Fractal dimension $D_{f}$ (filled diamonds; solid line is 
guide for eye) 
versus pinning strength $f_{p}$, 
and the fraction $N_{a}/N_{v}$ (open triangles; dotted line is guide
for eye)
of vortices moving inside the sample versus $f_{p}$.  
Both show a slope change near $f_{p} \approx 0.6f_{0}$.  
Lower left inset: Average distance between channels $d_{\rm perp}$
versus $f_{p}$.
Upper right inset: Variation with time of the tortuosity, 
$\sigma_{\tau}$, versus $f_{p}$.  Note that $\sigma_{\tau}$
drastically increases near the region where 
the power $S_{0}$ [Fig. 3] peaks.
}
\label{fig:2}
\end{figure}

{\it Channel network.\/}---We divide the sample into
a $300 \times 300$ grid to identify the vortex channels.
When a vortex enters a grid element, the counter associated
with that grid element is incremented, defining a 
``channel transit'' field.  All grid elements that are visited at 
least once are considered part of the channel network.
We
calculate the average rate $\Gamma_{\rm av}$ at which vortices move through
each grid site, and construct a distribution of $\Gamma_{\rm av}$ over
all the grid sites to indicate how frequently
individual channels were traversed.
Figure~\ref{fig:1} shows channels and distributions $P(\Gamma_{\rm av})$  
from four samples with different pinning strengths $f_{p}$.
For weak pinning, 
$f_{p} \lesssim 0.2 f_{0}$ [Fig.~\ref{fig:1}(a)], 
the channels cover the entire sample area relatively uniformly.  
Many grid sites are visited by vortices at a low rate, so 
$P(\Gamma_{\rm av})$ peaks at small $\Gamma_{\rm av}$.
As $f_{p}$ increases,
islands of pinned vortices 
(shown in white in Fig.~\ref{fig:1}) form and grow.
At the same time, favored channels for vortex flow appear.
Grid elements inside channel sites are frequently traversed
by vortices, so $P(\Gamma_{\rm av})$ extends to higher rates
$\Gamma_{\rm av}$.
At higher pinning forces when there are only a small number of channels, 
the grid elements inside channels produce
a distinguishable increase in $P(\Gamma_{\rm av})$ 
for $\Gamma_{av} \gtrsim 5 \times 10^{-4}$. 
The average separation between channels $d_{\rm perp}$
increases roughly linearly with increasing pinning strength $f_{p}$
as the ratio $n_{v}/n_{p}$ increases
[Fig.~\ref{fig:1}(b--d) and inset of Fig.~\ref{fig:2}], until 
for $f_{p} \sim 0.66f_{0}$, the
channels are no longer connected in the 
transverse $y$--direction, as is clearly visible 
in Fig.~\ref{fig:1}(d).  
This breakup represents a transition in the
nature of the plastic flow, as we shall show below.

{\it Fractal dimension.\/}---To quantify the effect of pinning 
strength $f_{p}$ on
the fractal dimension $D_{f}$ of the vortex channel network,
we use a box-counting algorithm \cite{Ref14} to find $D_{f}$,
and plot the results in Fig.~\ref{fig:2}.
Here,
$D_{f} = -\lim_{\epsilon \rightarrow 0} \log{N(\epsilon)}/\log{\epsilon}$, 
where $N(\epsilon)$ is the number of boxes of side $\epsilon$ required
to cover all grid sites belonging to the channels.  
The dimension $D_{f} \approx 2$ for 
low pinning 
strengths, $f_{p} \lesssim 0.2 f_{0}$, when vortices 
are flowing throughout the entire sample [Fig.~\ref{fig:1}(a)].
As $f_{p}$
is increased and the channel structure becomes more sparse
[Fig.~\ref{fig:1}(b--c)],
the fractal dimension $D_{f}$ decreases.  

Our samples with strong disorder have fractal dimensions close
to those predicted recently for channel networks in systems where 
elastic interactions are unimportant \cite{Ref15,Ref17}, 
For example, our sample with $f_{p}=0.75f_{0}$ has a 
fractal dimension of $D_{f} = 1.37$, close to
the mean-field prediction of 4/3 found in Ref.~\cite{Ref15} and
the value $1.38$ observed in Ref.~\cite{Ref16}.
At the strong pinning case of $f_{p} = 0.83f_{0}$, where there are
a few isolated channels in the sample, we find $D_{f} = 1.27$.
This is in 
reasonable 

\hspace{-13 pt}
agreement with simulations 
of non-interacting particles 
\cite{Ref15},
$D_{f} = 1.21$, and with other theories \cite{Ref17}, $D_{f} = 1.22$.
At the very highest pinning strength, $f_{p}=0.9f_{0}$,
only a single river passes through 
the sample, giving an extremely
low fractal dimension $D_{f} = 1.15$, in agreement with
the fractal dimension of the main channel of physical rivers 
\cite{Ref7}, 
$D_{f} = 1.14$---$1.20$.

The fractal dimension gives a static picture of the vortex
channels.  We probe the dynamics of the channels by considering
the fraction $N_{a}/N_{v}$ of vortices that move
a distance greater than the pinning diameter $2\xi_{p}$.
We find that changes in $N_{a}/N_{v}$ closely follow changes
in the 

\hspace{-13 pt}
fractal dimension $D_{f}$.
At low pinning strengths, 
$f_{p} < 0.2f_{0}$, 
all of the
vortices move, as seen in Figs.~\ref{fig:1}(a) 
and~\ref{fig:2}, 
indicating
mostly elastic motion.
The moving fraction $N_{a}/N_{v}$ decreases
with increasing pinning strength 
as the motion becomes plastic and some vortices remain permanently trapped 
in pinning sites.  
Fits to $N_{a}(f_{p})$ [dashed line in Fig.~\ref{fig:2}] and
$D_{f}(f_{p})$ [solid line]
indicate that there is
a small but noticeable change of slope in both quantities near
$f_{p}/f_{0} \sim 0.6f_{0}$.
As we shall see later, this occurs when the vortex channels change 
behavior from braiding $(f_{p}/f_{0} \lesssim 0.6)$ to
non-braiding $(f_{p}/f_{0} \gtrsim 0.6)$.

{\it Tortuosity and fluctuating braided channels.\/}---To 
examine the motion of individual vortices, we
compute the tortuosity of the path followed by each vortex.
The tortuosity $\tau$ measures the amount a path winds \cite{Ref7}:
$\tau = x/L$
where $x$ is the actual distance traveled by the vortex as it crosses
the sample, and $L$ is the width of the sample. 
Thus,  $\tau \geq 1$, and for a straight path, $\tau = 1$.  

\begin{figure}
\centerline{
\epsfxsize=3.4in
\epsfbox{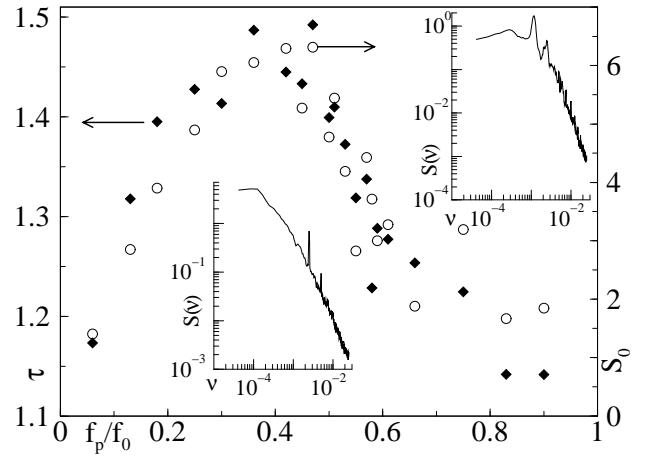}}
\caption{Tortuosity $\tau$ (filled diamonds) of vortex paths and
power $S_{0}$ in the second octave of the spectra (open circles), both
versus pinning force $f_{p}$. 
The same results hold for the third and fourth octaves.
A peak in the noise power was also observed in spectra obtained from
the voltage transverse to the driving direction.
Left inset: Spectra $S(\nu)$ for $f_{p} = 0.18f_{0}$.
Right inset: Spectra $S(\nu)$ for $f_{p} = 0.66f_{0}$.
The narrow band peaks in both spectra are caused by the regular
rate at which vortices are added to the sample,
while the peak at $\nu \sim 10^{-3}$ in the right inset is due to
the typical time of flight for motion through the isolated channels.
}
\label{fig:3}
\end{figure}

The plot of the average tortuosity $\tau$ shown in Fig.~\ref{fig:3}  
reveals a very interesting behavior that is not reflected in
the fractal dimension $D_{f}$.  For low pinning strengths
$f_{p}<0.05f_{0}$, $\tau \sim 1.1$, indicating that the vortex paths 
wind very little.
The tortuosity increases with pinning force 
as the pins become more effective and cause individual vortex paths to wind
around islands of pinned flux,
as in Fig.~\ref{fig:1}(b).
A vortex can follow a very tortuous
trajectory by crossing between what would have been
{\it distinct\/} paths at lower pinning strengths.  
The heavy crossing or braiding of channels
leads to a peak value of $\tau \approx 1.5$
for $f_{p} \approx 0.5f_{0}$.
As seen in the inset of Fig.~\ref{fig:2}, 
the variation in time of the tortuosity,
$\sigma_{\tau}$, also peaks near $f_{p} \approx 0.5f_{0}$.
For $f_{p} \lesssim 0.5 f_{0}$, the vortices follow a network of
{\it heavily} 

\hspace{-13 pt}
{\it braided\/} flow channels.
A remarkable drop in $\tau$ between 
$f_{p} \sim 0.5f_{0}$ and 
$f_{p} \sim 0.7 f_{0}$
to a saturated low value of 
$\tau \sim 1.15$ is clearly visible
in Fig.~\ref{fig:3}.
It is important to 
emphasize that 
the drop coincides with a transition 
from 
channels that are braided 
across the {\it entire\/} length of the sample at
intermediate pinning strengths, $0.2f_{0}<f_{p}<0.5f_{0}$
[Fig.~\ref{fig:1}(a--b)], 
to {\it isolated, unbraided\/} channels 
at higher pinning strengths, $f_{p}>0.6f_{0}$
[Fig.~\ref{fig:1}(d)], 
that are too far apart to significantly interact. 
This change is not merely a finite size
effect, since it is occurring over length scales significantly smaller
than the sample width.  

The crossover from increasing to decreasing tortuosity results
from the combined effects of simultaneously increasing the vortex-pin
interactions and the flux gradient.
Vortex-pin interactions are less
important at low pinning forces, 
and the vortices follow relatively straight paths.
As the pin strength increases, some vortices become trapped, the vortex paths
begin to wind, and the tortuosity increases.  
The flux gradient is also increasing, however, and
at the crossover point, the flux gradient begins
to dominate.
The vortices then flow directly down the steeper gradient, 
decreasing the tortuosity.  
We have observed the transition in samples of different $x$ direction
lengths: $18\lambda \times 36\lambda$, $36\lambda \times 36\lambda$,
and $48\lambda \times 36\lambda$.  The pinning force at which the 
transition occurs shifts downwards slightly as sample length increases.
In very long samples it is thus possible for the channel flow phase to
dominate for a fixed current and to be detected by local Hall probes.
A more detailed account of the effects of sample size will appear elsewhere.

{\it Voltage Noise.\/}---We next link the transitions in the vortex 
channel structure
with experimentally accessible voltage noise signals.
We sum the forces in the $x$--direction along a strip of the 
sample $5 \lambda$ in width to obtain our voltage signal.  We find the
spectrum of the resulting signal, $V_{x} = \sum_{i} {\bf f}_{x}^{(i)}$,  for
each sample configuration, and indicate the spectral power
by plotting the integrated noise power in one frequency octave 
$S_{0} = \int_{\nu_{0}}^{\nu_{1}} d\nu S(\nu)$
versus pinning strength $f_{p}$ in Fig.~\ref{fig:3}.  
Here, $\nu_{0} = 1.2 \times 10^{-4}$ and $\nu_{1} = 2.4 \times 10^{-4}$.
Remarkably, the form of $S_{0}$ {\it closely follows the tortuosity\/} 
$\tau$.  
This is because both measures are sensitive to the number of
metastable states accessible to the system.  
When $\tau$ is high, the vortices wander significantly in
the transverse direction, sampling many metastable states
and thus producing a high noise power.
The overall drop in noise power for $f_{p} \gtrsim 0.5f_{0}$
occurs as the amount of braiding in the channels decreases.
We can compare our results to experiments \cite{Ref10,Ref11}, in which
a peak in the noise power occurs near the depinning transition 
when plastic flow occurs and high $\tau$ is expected.
At higher currents, the pinning effectively becomes weaker, 
the vortices flow more elastically, $\tau$ is expected to be lower,
and a drop in noise power is observed.
This agrees with the results shown Fig.~\ref{fig:3}:
for the most plastic flow, the highest $\tau$ occurs and
the noise power is highest, while for weaker pinning, 
the vortices flow in straighter paths, $\tau$ is lower, and
the noise power drops.
Noise measurements can thus be used to probe the tortuosity $\tau$.

The shape of the noise curve changes significantly
at $f_{p} \sim 0.5f_{0}$ [insets of Fig.~\ref{fig:3}].
For $f_{p} \lesssim 0.5f_{0}$, 
the spectrum for $\nu \gtrsim 10^{-3}$
is of the form $S(\nu) \sim \nu^{-\alpha}$
[left inset of Fig.~\ref{fig:3}],
where $\alpha$ decreases for higher pinning strengths.
For $f_{p}<0.1f_{0}$, $\alpha \approx 2$, while for 
$0.2f_{0}<f_{p}<0.5f_{0}$, $\alpha \approx 1.7$, 
consistent with the experimental measurements of
D'Anna {\it et al.\/} \cite{Ref9}
and  Marley {\it et al.\/} \cite{Ref10}, respectively.  
During and after the drop in $S_{0}$, for $f_{p} \gtrsim 0.5f_{0}$, 
the spectrum $S(\nu)$ is no longer of a form that can 
be characterized by a unique slope.
Instead, the relatively straight, isolated channels produce
a time of flight signal in the spectrum similar to experimentally
observed signals \cite{Ref9}. 
No unique time of flight signal appears in our simulation 
at lower pinning forces, $f_{p} < 0.5f_{0}$, since
the many braided channels present 
lead to a spread in the tortuosities and a spread in the
time spent crossing the sample. 

{\it Summary.\/}---
Using novel measures of vortex channel structures, including
the fractal dimension of the channel network and the tortuosity
of individual vortex paths, we have provided strong evidence 
for a transition between two
distinct vortex plastic flow phases as a function of disorder strength.
We have shown that, as disorder increases,
the weak-pinning straight paths continuously 
evolve to a braided winding-channel
pattern 
that is characterized by high noise power and
a high tortuosity with large fluctuations in time.
A sharp drop in the tortuosity and noise power
for intermediate pinning signals a transition to flow in
non-braiding, isolated individual channels. 
The drop in noise power is consistent with experimental measurements 
\cite{Ref10,Ref11}.  
The transition is a universal property of the average pinning strength
and not a sample-dependent phenomenon, since the disorder 
configuration was different in every one of the 21
sample used.
This crossover among different dynamical flow regimes
as a function of pinning strength
may also be important to other slowly driven disordered systems such
as Wigner crystals, colloids, Josephson junctions, and magnetic
bubbles. 
Our observation that the tortuosity and noise power follow each other closely
is a novel result indicating that a macroscopic noise power measurement
gives direct insight into the microscopic tortuosity, a link that may 
be useful for systems with driven channels.
Our predictions can be tested through experiments such as
Lorentz microscopy or noise measurements in superconductors, 
and direct imaging of colloids. 

Computer services were provided by: the Maui High Performance Computing
Center, sponsored 
in part by grant F29601-93-2-0001;
and by the University of Michigan Center for Parallel Computing,
partially funded by NSF Grant No. CDA-92-14296.  CO was supported
by the NASA Graduate Researchers Program.


\begin{references}
\bibitem{Ref13} Our vortices are flux-gradient driven, and
thus no artificial uniform force is applied to them.
Refs.~\cite{Ref3-4,Ref5} consider uniform force cases.
\end{references}
\end{document}